# R&D progress on second-generation crystals for Laue lens applications

N. Barrière[α*], P. von Ballmoos[α], P. Bastie[β], P. Courtois[χ], N. V. Abrosimov[δ],

K. Andersen[χ], T. Buslaps[ε], T. Camus[α], H. Halloin[φ], M. Jentschel[χ], J. Knödlseder[α], G. Roudil[α],

D. Serre[γ], G. Skinner[η]

[α] Centre d'Etude Spatiale des Rayonnements, Université Paul Sabatier, 9, avenue du Colonel Roche, 31028 Toulouse – France

[β] Laboratoire de Spectrométrie Physique, 140, avenue de la Physique, 38402 Saint Martin d'Hères - France

[χ] Institut Laue Langevin, 6, rue Jules Horowitz - 38042 Grenoble - France

[δ] Institut für Kristallzüchtung, Max-Born-Strasse 2, D-12489 Berlin - Germany

[ε] ESRF, 6 rue Jules Horowitz, 38043 Grenoble - France

[φ] APC, Université paris 7 – Denis Diderot, 10 rue Alice Domon et Léonie Duquet – 75013 Paris - France

[γ] LA, Observatoire Midi-Pyrénées, 14 avenue Edouard Belin, 31400 Toulouse – France

[η] NASA-GSFC / University of Maryland and CESR, Toulouse - France

## ABSTRACT

The concept of a gamma-ray telescope based on a Laue lens offers the possibility to increase the sensitivity by more than an order of magnitude with respect to existing instruments. Laue lenses have been developed by our collaboration for several years : the main achievement of this R&D program was the CLAIRE lens prototype, which has successfully demonstrated the feasibility of the concept in astrophysical conditions. Since then, the endeavour has been oriented towards the development of efficient diffracting elements (crystal slabs) in order to increase both the effective area and the width of the energy bandpass focused, the aim being to step from a technological Laue lens to a scientifically exploitable lens. The latest mission concept featuring a gamma-ray lens is the European Gamma-Ray Imager (GRI) which intends to make use of the Laue lens to cover energies from 200 keV to 1300 keV.

Investigations of two promising materials, low mosaicity copper and gradient concentration silicon-germanium are presented in this paper. The measurements have been performed during three runs: 6 + 4 days at the European Synchrotron Radiation Facility (Grenoble, France), on beamline ID15A, using a 500 keV monochromatic beam, and 14 days on the GAMS 4 instrument of the Institute Laue Langevin (Grenoble, France) featuring a highly monochromatic beam of 517 keV. Despite it was not perfectly homogeneous, the presented copper crystal has exhibited peak reflectivity of 25 % in accordance with theoretical predictions, and a mosaicity around 26 arcsec, the ideal range for the realization of a Laue lens such as GRI. Silicon–germanium featuring a constant gradient have been measured for the very first time at 500 keV. Two samples showed a quite homogeneous reflectivity reaching 26%, which is far from the 48 % already observed in experimental crystals but a very encouraging beginning. The measured results have been used to estimate the performance of the GRI Laue lens design.

---

# 1. INTRODUCTION

The soft gamma-ray domain represents a unique window on the non-thermal Universe, allowing violent objects and matter under extreme physical conditions impossible to reproduce in the laboratory to be studied. But still nowadays, observing nuclear gamma-ray lines and continuum emissions from celestial sources remains very challenging because of the intense background count rate induced in detectors operated in space by cosmic rays, solar events, and van Allen belts. After the launch of the INTEGRAL mission by the European Space Agency (ESA), to keep improving the sensitivity of the future soft gamma-ray telescopes, it seemed mandatory to change radically the principle of the telescopes operating in this energy band. Thanks to efforts over several years on the development of Laue lenses, the focusing of gamma rays up to MeV energies is now enabled (as shown in ref. 1, 2), and ready to be used scientifically.

The Gamma-ray Imager mission (GRI) recently proposed to ESA in the framework of the Cosmic Vision 2015-2025 program intends to make use of a such a Laue lens to focus photons from 220 keV up to 1.3 MeV. A multilayer concentrator is also part of the mission for the coverage of energies between 30 keV and 300 keV. Since these focusing optics have focal lengths of 100 m, they are carried by a spacecraft flying in formation with another one carrying the focal plane instrument. In the proposed design the focal plane detector, which is common for both optics, is composed of a stack of planar pixelated CZT detectors, topped by a collimator and surrounded by more planar pixelated CZT detectors (see ref. 3 and 4 for a detailed description of the focal plane instrument, and ref. 5 for a complete overview of the GRI mission).

In the present paper, after having briefly recalled the principles at work in a Laue lens, we discuss design considerations of the GRI Laue lens and their consequences on the crystals requirements. Then we show that different kind of crystals, with mosaic and curved diffracting planes, can satisfy these requirements, and a comparison between them is made. Experimental results concerning copper mosaic crystals and silicon-germanium gradient crystals are presented, and finally the expected performance of the GRI Laue lens is derived from these results.

# 2. LAUE LENS PRINCIPLE

A Laue lens focuses gamma rays by using Bragg diffraction in the Laue geometry, which means that the rays go through the crystals. A large number of crystal slabs are arranged in concentric rings and orientated such that they diffract radiation coming from infinity toward a common point, the focus. In the simplest design each ring is composed of identical crystals and is symmetrical with respect to the line of sight of the lens. $\theta_B$, the angle of incidence of the rays with respect to the set of atomic planes of Miller index (hkl), is related to the diffracted energy $E$ at the $n$th order, by Bragg's law

$$2d_{hkl} \sin\theta_B = n\frac{hc}{E}, \qquad (1)$$

$d_{hkl}$ being the inter-plane distance, $h$ Planck constant and $c$ the velocity of light in vacuum.

Considering a focal distance $f$, the mean energy diffracted by a ring is thus given by its radius $r_i$, and the d-spacing of the crystalline planes used in the ring:

$$E_i = \frac{hc}{2d_{hkl}\sin\left(\frac{1}{2}\arctan\left(\frac{r_i}{f}\right)\right)} \approx 12.4 \frac{f}{d_{hkl}\,[\text{Å}]\,r_i} \text{ [keV]}. \qquad (2)$$

As we can see from the Bragg's law (1), perfect single crystals are not suitable for the realization of a Laue lens because they diffract almost only a single wavelength, behaving as a monochromator. In order to obtain an energy bandpass, crystals must present a range of atomic planes orientation. As discussed in the next section, there are two kinds of crystalline structures offering this mandatory feature: either mosaic structures or curved diffracting planes. In both case, the distribution of orientations are characterized by their full width at half maximum (FWHM) that is denoted $\Omega$ (called 'mosaicity' in the case of mosaic crystal, or angular spread in the general case). The bandpass of a crystal, and therefore of a full ring, is given by the following formula:

$$\Delta E \approx \frac{\Omega E_i}{\theta_B} \approx \frac{2\Omega E_i f}{r_i}. \qquad (3)$$

It is possible with this principle to cover a wide energy range by overlapping the bandpasses produced in different rings. This is obtained by using an identical d-spacing (same material, same set of diffracting planes) for several adjacent rings, so that the energy is shifted from one ring to the next of the amount given by $E_i - E_{i+1}$:

$$E_i - E_{i+1} \approx 12.4 \frac{f}{d_{hkl}[\text{Å}]} \frac{r_{i+1} - r_i}{r_i^2} = E_i \frac{r_{i+1} - r_i}{r_i} \quad [\text{keV}], \text{ with } r_i < r_{i+1}. \quad (4)$$

In the above formulae, the approximations are valid for small angles, which is the case with the energies of the soft gamma-ray domain.

### 3. THE GRI LAUE LENS DESIGN: CONSTRAINTS ON CRYSTALS

Table 1 summarizes the GRI Laue lens requirements that play directly on the design. The first point to take into account is the energy band, which comes straight from the scientific objectives of the mission[6]. The second point to be considered is the range of available radii for the crystals rings (table 2), limited on the inside by the multilayers mirrors, and on the outside by the diameter of the Soyouz rocket fairing[5]. With these constraints there is no focal distance that allows a single material to cover entirely in an acceptable way the energy bandpass required. For this reason the use of different diffracting materials (meaning different d-spacing) has to be envisaged. Obviously, broadening the choice of diffracting material offers more possibilities of arrangement and so, more chance to reach the optimum coverage.

| Item | Requirement | Goal |
| --- | --- | --- |
| Energy band(s) [keV] | 200 – 600<br>800 - 900 | 200 – 1300 |
| Continuum sensitivity, 3σ, 100 ks [ph cm$^{-2}$ s$^{-1}$ keV$^{-1}$] | 10$^{-7}$ | 3 10$^{-8}$ |
| Narrow line sensitivity, 3σ, 100 ks [ph cm$^{-2}$ s$^{-1}$] | 3 10$^{-6}$ | 10$^{-6}$ |
| Field of view [arc min.] | 5 | 10 |
| Angular resolution [arc sec.] | 60 | 30 |

**Table 1** : Goals and requirements for the performances of the GRI Laue lens.

| Parameter | Constraint |
| --- | --- |
| Radii range available for crystals [cm] | 60 < r < 180 |
| Circumference of a ring available for crystals [%] | 80 |
| Maximum crystals mass [kg] | 250 |

**Table 2** : Constraints on the collecting area of the Laue lens and on the crystals total mass.

### 3.1. Crystal requirements.

Selected materials have first and foremost to exist in a crystalline state and be reproducible with good production yield, and also be mechanically robust enough to allow accurate mounting. Secondly, the crystals have to have an angular spread in their diffracting planes, i.e. to have either a mosaic structure or a curvature in their diffracting planes. Since we consider entire rings of crystals focusing the same energy band, the lower limit on $\Omega$, the FWHM of the angular distribution of plane orientations, is given by geometrical considerations: In order that the bandpasses of adjacent rings overlap enough to produce a flat energy coverage, it is necessary that $E_i - E_{i+1} \leq \Delta E_i$. By combining equations (3) and (4), one obtains the minimum $\Omega$ value for the crystals as a function of the focal distance and the radial size of the crystal slabs:

$$\Omega \geq \frac{r_{i+1} - r_i}{2f}. \quad (5)$$

In the case of mosaic crystals (Gaussian-like energy bandpass), this condition corresponds to a continuous flat response at twice the peak effective area of a single ring. For mosaic crystals Ω could in principle be reduced by a factor of two from this value, but this would unduly tighten the constraint on the mounting precision.

The GRI Laue lens features a focal distance $f$=100 m, which appears to be the best compromise to achieve the energy coverage required, and the crystal cross section has been chosen to be 15 mm x 15 mm. This size is the result of a trade-off between the resulting focal spot size on one side, and on the other side, the alignment requirements and the total number of crystals (cut, orientation, mounting, collecting area lost in the inter-crystal spacing). Taking into account an additional space between rings of 0.5 mm, these values imply $\Omega \geq 16$ arcsec.

Knowing that a larger value of Ω decreases the ideal performance of the lens but relaxes the constraints on crystal mounting precision, thus making it easier to obtain the calculated performances, a compromise has been made with crystals featuring an angular spread of 30 arcsec.

## 4. DIFFRACTION THEORY IN CRYSTALS FOR LAUE LENS APPLICATION

### 4.1. Mosaic crystals

Diffraction in mosaic crystals is described by the Darwin model. These crystals are represented by an assembly of independent small perfect crystals, the crystallites, whose orientation have a Gaussian distribution around the mean orientation of the crystal. The FWHM of the crystallite angular distribution function is the mosaicity Ω. The peak reflectivity of a mosaic crystal, which is defined for a given wavelength by the ratio of the intensity in the incident beam to the maximum intensity in the diffracted beam, is given by the following formula (see ref. 7 for a complete review of the diffraction in mosaic crystals applied to a Laue lens):

$$r^{max} = \frac{I_h(T_0)}{I_0(0)} = \frac{1}{2}\left(1 - e^{-2\sigma T_0}\right) e^{-\frac{\mu T_0}{\cos\theta_B}}, \quad (6)$$

where $T_0$ is the crystal thickness, $\mu$ is the linear absorption coefficient, and $\sigma$ can be considered as a linear diffraction coefficient (derived from the dynamical theory of diffraction).

This model shows two important features: firstly, even without taking into account the absorption in the crystal, the diffraction efficiency is limited to 50 %, which is explained by the equilibrium between direct and diffracted beams occurring in the crystal. This maximum is reached in the case of an 'ideally imperfect' crystal, which means that the crystallites mean size is smaller than the extinction length in the crystal (energy dependent). Secondly, the shape of the energy bandpass is dominated by the crystallite angular distribution function, being therefore close to a Gaussian curve. Consequently, the spatial distribution of photons diffracted by a mosaic crystal is Gaussian-like as well (as shown in Fig. 1), which leads to an enlargement of the focal spot at long focal distances, the so-called 'mosaic defocusing' effect.

### 4.2. Crystals with Curved diffracting planes .

Diffraction in crystals with curved atomic planes is described by the Takagy–Taupin's theory of strongly distorted crystals[8], which has an analytical solution in the case of a constant strain gradient, i.e. a crystal with spherically curved planes. Under this condition, the peak reflectivity is given by the following formula:

$$r^{max} = \frac{I_h(T_0)}{I_0(0)} = \left(1 - e^{-\frac{\pi^2}{2\alpha}}\right) e^{-\frac{\mu T_0}{\cos\theta_B}}, \quad \alpha > 1, \quad (7)$$

α being defined as the product of the planes curvature and the extinction length divided by the Darwin width, which is proportional to the curvature of the diffracting planes multiplied by the square of the energy: $\alpha = \frac{\Omega/T_0 \, \Lambda}{\delta} \propto \frac{\Omega E^2}{T_0}$.

We see immediately the first interest of these crystals: their diffraction efficiency is not limited to 50%. Disregarding the absorption in the crystal, the diffraction efficiency can reach 100 %. Another valuable point is the top hat shaped energy bandpass that implies a well-defined spatial distribution of the diffracted photons (as shown in Fig. 1). But we also see that the reflectivity will decrease quickly with the angular spread of the crystal and even faster with the energy, meaning that these crystals will no longer be interesting for large bandpass at high energy.

There are 3 ways to obtain curved crystalline planes; the two first consists in applying either a thermal gradient or a pure mechanical bending to a perfect single crystal[9]. The third, which has been developed in the

Institute of Crystal growth (IKZ, Berlin, Germany), consists in growing a binary alloy in which the relative concentration varies along the growth axis, inducing a curvature of the planes perpendicular to the concentration gradient axis. This method is preferred for a space borne Laue lens application since the crystal slabs obtained can feature the curvature intrinsically without any external device.

In the case of a $Si_{1-x}Ge_x$ concentration gradient crystal, the angular spread of the diffracting planes $\Omega$ is related to the Ge concentration $C_{Ge}$ gradient through the crystal thickness $T_0$:

$$\nabla C_{Ge} \approx 25 \frac{\Omega}{T_0}. \qquad (8)$$

Thus in order to extract homogeneous crystal slabs from an ingot, one has to grow a crystal with a constant gradient of concentration, which has been done recently for the first time.

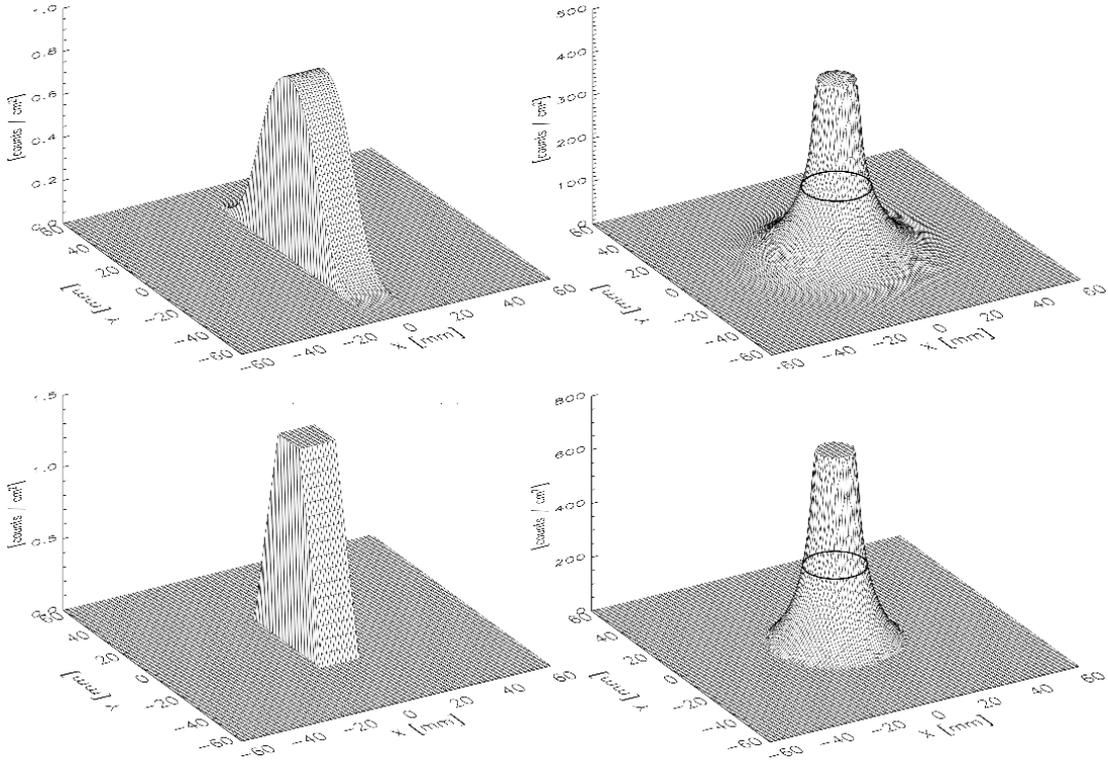

**Figure 1 :** Simulations of the spatial distribution of photons focused by a crystal ring, 100m away from the focal plane. In all cases crystals are 15 mm x 15 mm, have an angular spread of 30 arcsec, and focus an energy band centred on 300 keV. On the left panel the ring contains only 1 crystal, while on the right panel the ring is made of 500 crystals. The black line on the right panel shows the radius on the detector that optimises the detection significance in the presence of a detector background. In the upper panel, the simulated rings have a radius of 1.980 m and the crystals are Cu(111) of 2.7 mm thickness. On the lower panel, the ring radius equals 1.318 m, and the crystals are $Si_{1-x}Ge_x$ gradient (111) of 24.4 mm thickness.

## 4.3. Comparison

Figure 1 shows a comparison between the focal spot produced at 100 m by Cu(111) mosaic crystals and by SiGe(111) gradient crystals. On the right panel the 'lens' is made of a single ring containing 500 crystal slabs of 15 x15 mm with an angular spread of 30 arcsec, and diffracting an energy band centred on 300 keV. The simulated source is a point source on-axis giving a constant flux of 1 ph/cm$^2$/keV between 270 and 330 keV (larger than the bandwidth of the rings). The detection significance calculated according the following formula is used to compare the two focal spots: $\sigma = \frac{S}{\sqrt{S+N}} \approx \frac{S}{\sqrt{N}} \propto \frac{S}{\sqrt{V_{det}}} \propto \frac{S}{r_{focus}}$, $S$ being the signal count, $N$ the background count, $V_{det}$ the useful volume of the detector assuming a cylinder, and $r_{focus}$ the optimal radius of this cylinder.

As shown in the Table 3, the SiGe crystals have the potential to increase by about 70 % the detection significance of the Laue lens. But one has to take care with this result: as we also see in the table 3, this performance has a cost in terms of mass: in this particular case the mass is increased by 131% compared with Cu crystals. Moreover, as we will see in the next section, the gradient crystals do not always reach their theoretical maximum of diffraction efficiency, which can temper the gain in sensitivity. However the above calculation shows that these crystals are potentially of interest for Laue lenses and their development is justified.

|      | Thickness [mm] | Ring radius [m] | Mass of crystals [kg] | $r_{focus}$ [mm] | Counts encircled | Diffr. Photons encircled [%] | $\frac{S}{r_{focus}}$ [cts/mm] |
|------|------|------|------|------|------|------|------|
| Cu   | 2.7  | 1,98 | 2.74 | 12.9 | 1451 | 53.7 | 112 |
| SiGe | 24.4 | 1,318 | 6.34 | 11.2 | 2146 | 64.9 | 192 |

**Table 3** : Details and analysis of the comparative simulations (see figure 1 and the text).

## 5. CRYSTAL MEASUREMENTS

Since the end of the program CLAIRE, many experiments have been set-up to measure and develop crystals that could enhance the overall focusing efficiency of Laue lenses[10]. Among them, low mosaicity copper[11] produced by the monochromator group of the Institute Laue-Langevin (ILL, Grenpble, France) and $Si_{1-x}Ge_x$ gradient crystals[12] grown in the Institute of Crystal Growth (IKZ, Berlin, Germany) have been the most extensively studied, but they are not the only option for Laue lens applications; Mosaic germanium crystals[13], as used on the CLAIRE lens prototype[1,2] remain interesting.

Various crystal samples have been measured during three runs: 6 + 4 days at the European Synchrotron Radiation Facility (ESRF, Grenoble, France) using 300 keV, 500 keV, 700 keV beams, on beamline ID15A[14,15], and 14 days on the GAMS 4 instrument of the ILL featuring a highly monochromatic beam of 517 keV and 815 keV photons. In the present paper, we will only describe the results obtained around 500 keV with Cu and SiGe crystals.

The crystal samples have been investigated by measuring rocking curves (RC); The measurement involves rotating the sample in a quasi-parallel and monochromatic beam in order to plot the diffracted intensity as a function of the angle of incidence. This measurement can either be done with the detector in the direction of the transmitted beam (transmission RC), so the diffraction removes intensity from the transmitted beam, or with the detector in the diffracted beam (diffraction RC). In either case, the RC gives the angular distribution of the diffracting planes;its FWHM is a direct measurement of the mosaicity $\Omega$ (or angular spread). The transmission RC allows in addition to get the diffraction efficiency disregarding the absorption through the crystal, since the diffracted counts are removed from the intensity recorded when no diffraction occurs. The reflectivity of the sample can then be determined by multiplying the diffraction efficiency to the attenuation factor due to linear absorption in the sample (transmission coefficient).

### 5.1. Low mosaicity copper crystals

Up to a couple of years ago, the main problem with copper crystals was their too large mosaicity, ranging around a few minutes of arc at best. This problem has been overcome in ILL by Courtois et al. during summer 2005 when a refined growth process allowed large ingots (12 kg) of very high quality to be obtained[11]. The sample used here comes from one of these new ingot, number 805. The aim of these measurements was to investigate the mosaicity range, the agreement with the Darwin model and the homogeneity in the sample and between samples. The sample presented in figure 2 (Cu 805.23) was measured at 6 spots at 489 keV. One of the best rocking curves obtained is shown on the left hand side of figure 2. We can see that it is perfect for the GRI Laue lens: the diffraction efficiency reaches 48.5%, very close to the theoretical maximum. The mosaicity (which is directly given by the FWHM of the rocking curve) is 29.8 ± 8 arcsec, and the rocking curve is in good accordance with the Darwin model. Taking into account the absorption in the sample, the reflectivity reaches 25%, which is very satisfactory for the (220) reflection. Another aim of measuring crystals is to extract the parameters useful as input in our lens simulations. For mosaic crystal the mean crystallite size plays an important role in the diffraction efficiency. In this example, it is 100 μm (to be compared with the extinction length of 240 μm at 489 keV).

Despite this good example proving that our expectations in terms of crystals quality for the GRI Laue lens are possible, this sample does not have a perfect homogeneity. Figure 3 shows the mosaicity measured along the 6 spots with a standard deviation of 6.5 arcsec around the mean of 25.4 arcsec. This dispersion is believed to be due to the origin of this sample that comes from the neck of the ingot, a transition region between the crystal seed presenting a mosaicity of several minutes of arc and the bulk which is expected to be of high quality and uniformity. Much more homogeneous crystals are expected from the bulk of the ingot but so far it has not been possible to measure other samples coming from the same ingot. Such measurements are scheduled before summer 2008, they should demonstrate that copper is a usable material for the realisation of a Laue lens.

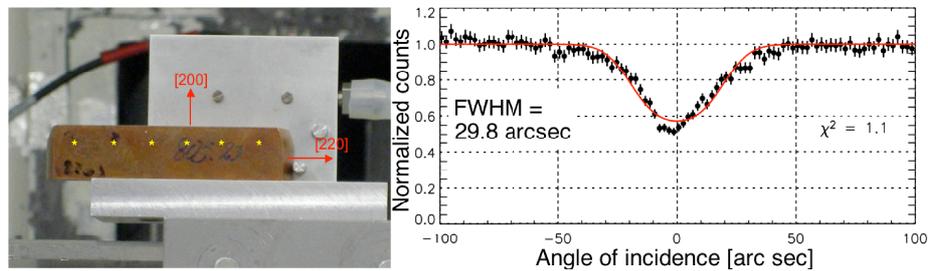

**Figure 2 :** Copper sample 805.23 grown, cut and etched in ILL, Grenoble, France. On the right is shown one of the best rocking curves recorded on this sample, investigated on its (220) planes. These measurements have been performed at the ESRF, using a 489 keV beam of 2.5mm x 0.7 mm cross-section with a divergence of 2.5 arcsec. The thickness of the sample is 9 mm, so the transmission coefficient at 489 keV is 0.52.

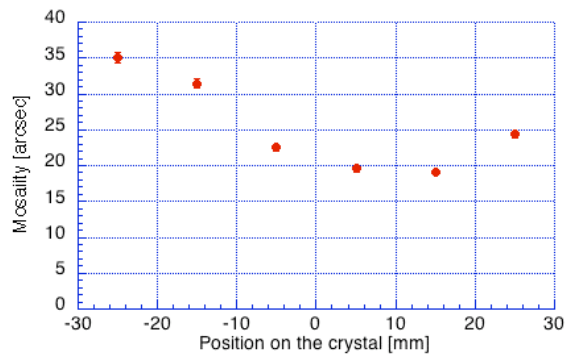

**Figure 3 :** Mosaicity along the 6 measurement spots of the sample presented in Figure 2. The mosaicity ranges between 18 and 35 arcsec, with a mean and a standard deviation of 25.4 ± 6.5 arcsec. This dispersion, which is quite significant, is believed to be due to the origin of the sample that was cut from the neck of the ingot.

### 5.2. SiGe gradient crystals

As explained in section 4, crystals with curved diffracting planes are of great interest for use in Laue lenses. But in order to be usable they have to be homogeneous over their collecting face. When a SiGe ingot is grown, the Ge concentration naturally evolves according to the Scheil-Pfann expression[12] which induces an increasing concentration gradient along the growth axis and consequently a variation of the curvature of the planes which translate into a variation of angular spread (see Equation 8).

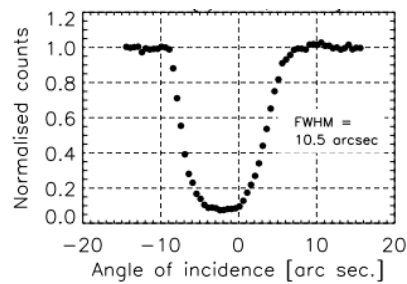

**Figure 4 :** Transmission rocking curve of the varying-gradient sample SiGe-10.3 measured at 297 keV in ESRF using a beam of 2.0 mm x 0.5 mm diverging with an angle of 1.8 arcsec. The sample measures 20 mm (gradient direction) x 27 mm x 20mm. This rocking curve is made for (111) diffracting planes through 20 mm, which gives a transmission coefficient equal to 0.61. Finally, the reflectivity reaches 59% in a bandpass of 10.5 arcsec at 297 keV, which are the best values ever seen.

We first studied varying gradient crystals in order to check the relation between the diffraction efficiency and the α parameter. The data obtained on a variety of samples are being processed and should be published soon.

But these measurements have shown that SiGe gradient crystals have truly a great potential, as shown in Figure 4. This RC, which was obtained at the ESRF in November 2006 using a beam of 297 keV, presents a diffraction efficiency reaching 97%, which gives to the sample a reflectivity of 59% with an angular spread of 10.5 arcsec - an amazing result! But, since this sample had a varying gradient, only a small area diffracts with such a good efficiency.

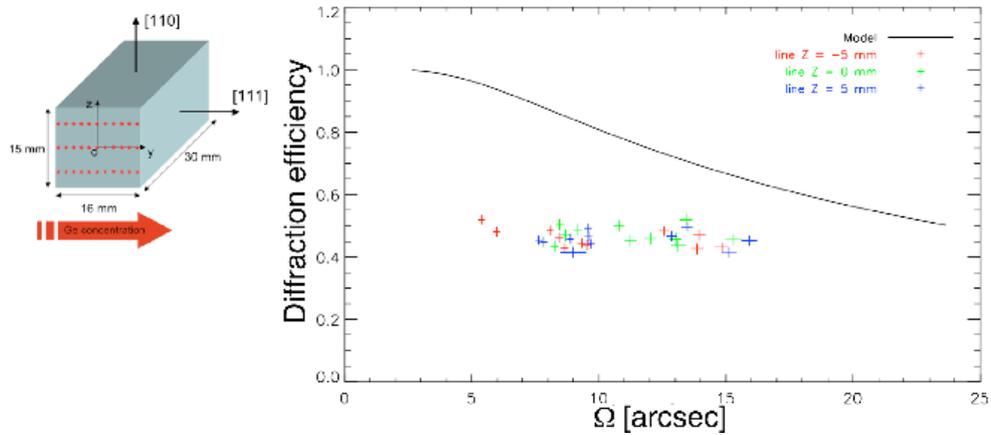

**Figure 5 :** Left: Sketch of the sample SiGe-312 (15mm x 16mm x 30mm) investigated at ESRF using a 0.8mm x 0.2mm beam of 0.72 arcsec divergence at 495 keV (the arrow symbolizes the Ge concentration gradient). 3 lines of 11 spots have been measured using the (111) reflection. On the right hand side is plotted the peak diffraction efficiency (disregarding the absorption through the crystal) VS the angular spread $\Omega$. The black line shows the theoretical value taken from equation (7).

The second step was to produce crystals with a constant gradient value optimized for a given energy. We have chosen 500 keV, which is thought to be the highest energy where SiGe gradient crystals can be interesting (i.e. better than copper mosaic crystals). Two experimental ingots from which only one crystal slab featuring a constant Ge concentration gradient could be extracted have been produced at IKZ (by growing ingots having a specific shape, N. Abrosimov, private com.).

Results obtained on samples SiGe-312 and SiGe-322 measured respectively at the ESRF using a beam of 495 keV and at ILL using a beam of 517 keV are shown in Figures 5 and 6. In both cases, the reflectivity and the homogeneity of samples have been investigated in many spots. Both samples gave very similar results, with relatively homogeneous diffraction efficiencies around 48 %. Detailed results are shown in Table 4. The achieved reflectivities are 25 % and 26%, which is good, but far from the theoretical maximum considering the mosaicity range observed (see Fig. 5 and 6). On the other hand, the homogeneity of the curvature is still not completely satisfying, especially in the sample SiGe-312 where the amplitude of the variation is as large as 11 arcsec. Thus these attempts have mixed results, with a reflectivity as good as the one obtained with copper mosaic crystals, but also with a (too) small angular spread having relatively large variations.

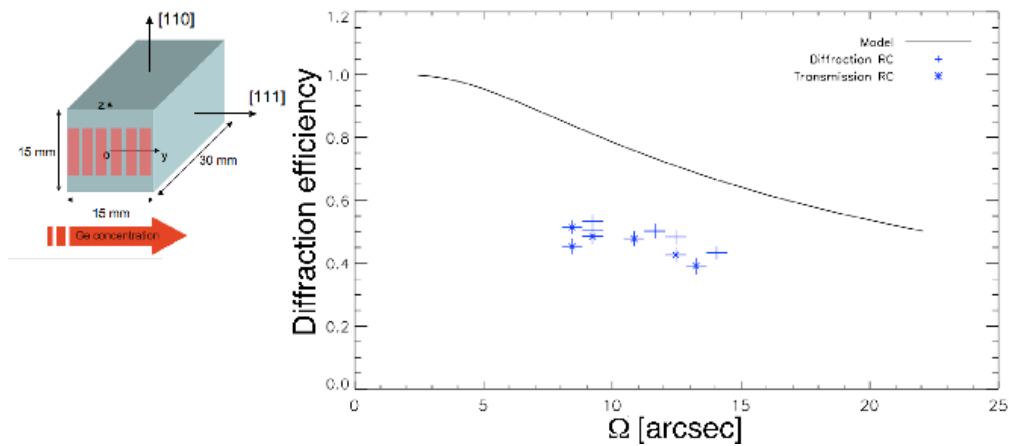

**Figure 6 :** Left : Sketch of the sample SiGe-322 (15mm x 15mm x 30mm) investigated at ILL using a 7.6mm x 1.7mm beam of 2 arcsec divergence at 517 keV. 6 spots have been done using the (111) reflection, through a thickness of 30 mm. Right: same plot

as in the figure 5 but measured on the sample SiGe-322. The plus signs represent data from diffraction RC while the asterisks represent data obtained from transmission RC.

As can be seen in the plots (Fig. 5 and 6.), the theoretical maximum diffraction efficiency drops quite fast with the angular spread. Since we are looking for crystals in the range of 30 arcsec, 500 keV appears to be the upper limit for SiGe crystals. That is why these crystals are only used for energies up to 330 keV in the proposed GRI Laue lens design. Using the present experimental results, a new iteration will be made with IKZ with the aim of enhancing the homogeneity of the curvature in crystals optimized for 300 keV.

| sample | Peak diffraction efficiency | Ω [arcsec] | Peak reflectivity |
|---|---|---|---|
| SiGe-312 | 0.46 ± 0.03 | 11 ± 2 | 0.25 ± 0.02 |
| SiGe-322 | 0.47 ± 0.04 | 11 ± 2 | 0.26 ± 0.02 |

**Table 4 :** Results of the measurements done on the two constant-gradient SiGe samples. The values quoted are means, while the uncertainty is the standard deviation about the mean. The reflectivity is calculated taking into account the absorption in the 30 mm thickness of the samples.

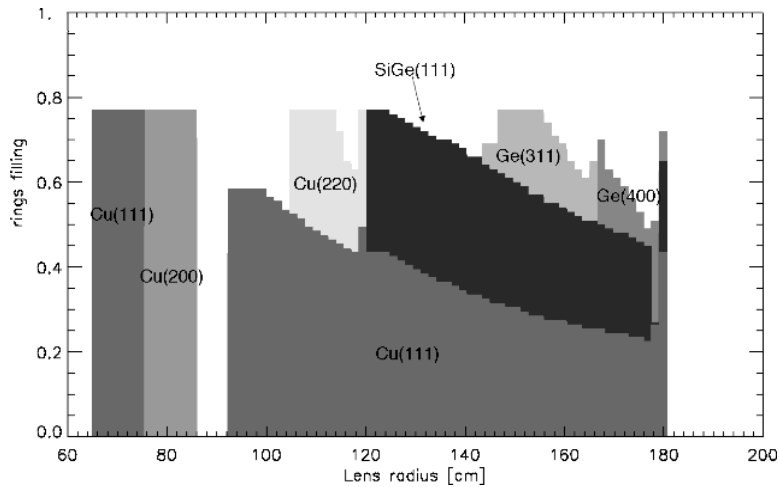

**Figure 7 :** Crystal distribution represented as the fraction of each ring filled with crystals, as a function of ring radius. A single ring can be shared between many different materials in order to achieve the coverage of the required energy bandpasses. The lens structure is radially shared in 8 petals covering 282°, meaning that 80 % of a ring circumference can be covered by crystals[5].

## 6. THE GAMMA-RAY IMAGER LAUE LENS

The proposed design, with a focal length of 100 m, makes use of three different materials and six different sets of diffracting planes to cover two energy bands (with the first order diffraction) from 220 keV to 640 keV and from 800 keV to 900keV as required for the mission (table 1). The materials are Cu, Ge and SiGe, but there is an option to replace the SiGe by Ge crystals if the promise of the SiGe does not materialise.

In the current mission design, the Laue lens crystals are mounted on 8 petals, each having an opening angle of 36.4° and a length of 121 cm [5] meaning that 80% of the area between 59 cm and 180 cm of radius is available for crystals. To design the lens, a new code has been added to lens simulation tools already developed, which allow combining in a more complex way different materials and diffracting planes to get an effective area meeting the requirements. The simulation of mosaic crystals (Cu and Ge) is based on the Darwin model using the dynamical theory of diffraction, using in both cases a crystallite size of 100 μm as obtained from measurements. The SiGe crystals are modelled using equation 7, which could seem optimistic regarding our last experiment, but which fit very well past measurements performed at lower energies[16].

The crystal distribution as function of the radius of the lens is shown in figure 7, while the resulting effective area is shown in Figure 8. The high-energy band, is covered by the inner rings (Cu(111) and Cu(200)) and is reinforced by Cu(220) crystals. Energies between 220 keV and 330 keV are covered by SiGe(111) crystals. Cu(111), Ge(311) and Ge(400) overlap with each other to cover energies from 330 keV up to 640 keV. The present configuration is made of 28000 crystals slabs of 15 mm x 15 mm, with angular spread ranging between 30 and 40 arcsec, representing a total mass of 252 kg.

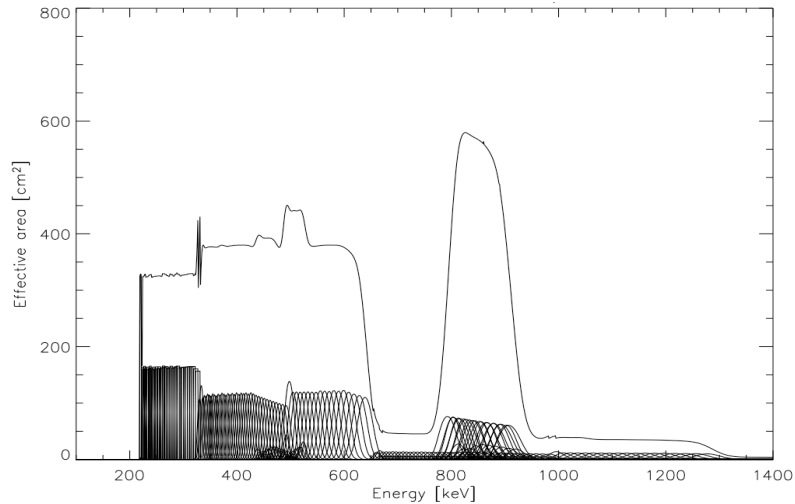

**Figure 8 :** Effective area of the GRI Laue lens made of copper, germanium, and silicon-germanium crystals. A thickness of 1.5 mm of aluminium is taken into account to simulate the structure absorption.

## CONCLUSION

The performance of a future Laue lens depends mainly on the diffraction properties of its individual crystals. Despite the fact that mosaic crystals produce a focal spot not ideally concentrated on the detection plane, the copper crystal that we measured showed excellent features with peak reflectivity up to 25 % at 500 keV and a mosaicity of 30 arcsec. Even with the problem of homogeneity - that is probably already solved (though measurements are not yet available) - copper crystals represent a good material to build an efficient Laue lens.

We have also shown that Laue lenses can benefit from crystals having curved diffracting planes. Composition gradient silicon-germanium crystals are currently under development and already show a great potential for low energies (up to 400 keV seem reasonable). Coming experiments will investigate the possibility of producing constant gradient crystals with larger angular spread, and with a better yield (many samples cut in each ingot). On the other hand, new binary alloys that could produce curved diffracting planes efficient at higher energies (e.g. germanium-tin) have also begun to be studied. These crystals developments are currently done in the framework of two R&D programs: one is funded by the French Space Agency CNES and the second is funded by ESA.

Based on present crystals performance, a Laue lens design for the GRI mission have been made. This study, which takes into account the technical constraints coming the carrying satellite, clearly shows that a realistic Laue lens would be feasible, and would allow a future space borne gamma-ray observatory such as GRI to make a sensitivity leap of at least one order of magnitude compared with existing missions.

## ACKNOWLEDGMENTS

The authors acknowledge continuing support from the French Space Agency CNES.